\documentclass[10pt,conference,hidelinks]{IEEEtran}

\usepackage{url}
\usepackage{subfig}
\usepackage{graphicx}
\usepackage{amsmath,amssymb}
\usepackage{hyperref}
\usepackage{multirow}
\usepackage{listings}
\usepackage{tabularx}
\usepackage{xcolor}
\usepackage{comment}
\usepackage{listings,newtxtt}
\usepackage{tikz}
\usepackage{xcolor}

\usepackage{pbox}
\usepackage{comment}
\usepackage{csvsimple}
\usepackage{fancyhdr}
\usepackage{mathtools}
\usepackage{nicefrac}
\usepackage{caption}
\usepackage{amsmath}
\usepackage{makecell}
\usepackage{pifont}
\usepackage{tikz}

\usepackage{siunitx}
\begin{document}

\newcommand{\gem}{gem{5}}
\newcommand{\bear}{BEAR-Wr-Opt}
\newcommand{\oracle}{Oracle}

\title{Enabling Design Space Exploration of DRAM Caches in Emerging Memory Systems}

\author{
\\\IEEEauthorblockN{Maryam Babaie, Ayaz Akram, Jason Lowe-Power}
\IEEEauthorblockA{Department of Computer Science,
University of California, Davis\\
Email: \{mbabaie, yazakram, jlowepower\}@ucdavis.edu}}

\maketitle
\thispagestyle{plain}
\pagestyle{plain}

\begin{abstract}
The increasing growth of applications' memory capacity and performance demands has led the CPU vendors to deploy heterogeneous memory systems either within a single system or via disaggregation.
For instance, systems like Intel's Knights Landing and Sapphire Rapids can be configured to use high bandwidth memory as a cache to main memory.
While there is significant research investigating the designs of DRAM caches, there has been little research investigating DRAM caches from a full system point of view, because
there is not a suitable model available to the community to accurately study large-scale systems with DRAM caches at a cycle-level.
In this work we describe a new cycle-level DRAM cache model in the \gem{} simulator which can be used for heterogeneous and disaggregated systems.
We believe this model enables the community to perform a design space exploration for future generation of memory systems supporting DRAM caches.

\end{abstract}


\section{Introduction}

Today's computing systems must meet the large processing power and memory capacity demands of a broad range of applications including machine learning, artificial intelligence, graph analytics, etc.
These applications demand both large capacity and high performance from their main memory technology.
Unfortunately, memory technology is not a panacea and high-performance memory (e.g., HBM) comes with low capacity and high-capacity memory (e.g., NVRAM) comes at the cost of lower performance.
Current high bandwidth memory technologies provide limited capacities such as HBM. Other memory technologies with large capacity usually have limited bandwidth and high latency, such as non-volatile memories (NVRAM).
Thus, to provide both high-performance and high-capacity memories to fulfill applications' memory requirements, vendors have moved to heterogeneous memory systems where the processing units can use multiple memory devices.
This trend is accelerating with new interconnect technologies such as Compute Express Link (CXL) which allows systems to have both local memory and remotely attached memory devices.
For instance, Intel's Sapphire Rapids will provide HBM2, DDR5, 3DXPoint, and CXL support within the same system.
Another trend to respond to applications' memory demand is disaggregated memory systems where they employ different memories and interconnects in their local or remote nodes.
These kinds of disaggregated memory systems with local and CXL-attached memories are anticipated to be highly employed in cloud services~\cite{lipond}.

One way to implement these heterogeneous memory systems without burdening the programmer with manual data movement is to use the faster memory as a cache of the slower memory.
Hardware-managed DRAM caches are seen as one way to enable heterogeneity and disaggregation in memory systems to be more easily programmable.
These \emph{DRAM-based} caches can hide the latency of lower bandwidth or higher latency memories in heterogeneous systems.
The last decade has seen significant academic research on DRAM caches, and today these ideas are becoming a reality with CPU vendors implementing DRAM cache-based computer systems, e.g., Intel's Cascade Lake and Sapphire Rapids~\cite{sapphire}.

Despite the large body of research on DRAM caches, recent work has shown that these transparent hardware-based data movement designs are much less efficient than manual data movement~\cite{hildebrand2021case}. While the recent work investigating Intel's Cascade Lake systems provides some insight into real implementations on DRAM caches~\cite{hildebrand2021case,izraelevitz2019basic,wang2020characterizing}, there is a gap in the community's access to cycle-level simulation models for DRAM caches for further studies.

This paper describes a new \gem{}-based DRAM cache model capable of modeling different topologies of DRAM cache designs. We mainly enable design space exploration of a unified (inspired from Intel Cascade Lake) or a disaggregated (useful for heterogeneous and disaggregated memory systems) DRAM cache hardware.

Previous work has explored many aspects of DRAM cache design in simulation such as the replacement policy, caching granularity~\cite{qureshi2012fundamental,jevdjic2013stacked}, DRAM cache tag placement~\cite{huang2014atcache,loh2012supporting,loh2011efficiently}, associativity~\cite{qureshi2012fundamental,kotra2018chameleon,young2018accord}, and other metadata to improve performance~\cite{loh2011efficiently,jevdjic2013stacked,young2018accord}.
Trace-based or non-cycle-level simulation can appropriately evaluate these mostly high-level memory system design investigations. However, as shown in recent work, the micro-architecture of the DRAM cache interface and controller can lead to unexpected performance pathologies not captured in these trace-based simulation studies. For instance, a dirty miss to the DRAM cache requires up to \emph{five accesses} to memory in Intel’s Cascade Lake~\cite{hildebrand2021case}.
This interference between extra accesses to the DRAM to fulfill a single demand access causes performance degradation not seen in prior simulation studies~\cite{hildebrand2021case}.

To better understand these realistic DRAM cache systems, it is imperative to build a detailed DRAM cache simulation model to perform a design space exploration of DRAM caches for emerging memory systems. The previous research works on DRAM cache design improvements do not provide any (open source) DRAM cache modeling platform for the community to perform a detailed micro-architectural and timing analysis. In this work we describe a cycle-level model of DRAM caches for heterogeneous and disaggregated memory systems in the \gem{} simulator~\cite{lowepower2020gem5}.

Our model is capable of implementing different caching policies and architectures. The baseline protocol of our model takes inspiration from the actual hardware providing DRAM cache in Intel's Cascade Lake which has direct-mapped, insert-on-miss, and write-back caching policy, and tags and metadata are stored in the ECC bits along with the data in the same cache location. Our model leverages the cycle-level memory models of \gem{} which include DRAM~\cite{hansson2014simulating} and NVRAM~\cite{gem5-workshop-presentation} models in \gem{}. Our model implements the timing and micro-architectural details enforced by the memory interfaces including the memory technology timing constraints, scheduling policies, and internal buffers. This enables a cycle-level full system analysis of the DRAM caches in emerging memory systems, which is not possible by the prior works. To demonstrate such capabilities, we investigate three case studies.

First, we answer the question of \emph{how a DRAM cache system's performance compares to those without a DRAM cache.}
We find out for the baseline DRAM cache architecture (similar to Intel's Cascade Lake), the system with the DRAM cache has a lower throughput than the same system without a DRAM cache.
We show similar results to the DRAM cache hardware studies. We demonstrate that the interference between the demand accesses and the extra accesses to the DRAM cache and main memory devices to respond to the demand causes bandwidth under-utilization for both devices.

In the second case study, we investigate optimized DRAM cache designs to mitigate the overheads observed by the baseline DRAM cache architecture in the first case study.
First, we implement the technique proposed by BEAR DRAM cache~\cite{chou2015bear} on top of the baseline design which optimizes the write demands that hit on DRAM cache.
Second, on top of the first optimization, we implement another optimized design which
optimizes write hits as well as all demands that miss on the DRAM cache to a clean cache line. We show that both designs provide better throughput compared to the baseline
architecture and the second optimized design outperforms the first one. However, a system without a
DRAM cache still outperforms both optimized designs.

In the third case study, we answer the question \emph{what is the impact of latency of the link between local DRAM caches and remote backing stores on the
performance of systems?} For this study we consider the baseline DRAM cache design, once backed up by a DDR4 far memory, then by a NVRAM far memory.
We find that as the far memory link latency increases, the system throughput drops.
However, with increased link latency, the performance of the NVM system drops less than the DDR4 system.
Moreover, we find out once the link latency of the backing store increases, DRAM cache finally outperforms the same systems without a DRAM cache.

We have made the DRAM cache model presented in this work open-source and publicly available to the research community~\cite{dcacheGem5Code}. This work will be integrated into \gem{} mainstream. We believe this model enables community to perform research for future heterogeneous and disaggregated memory systems supporting DRAM caches.

\section{Design of DRAM cache model} \label{design}

We strive to support a generic DRAM cache model. For this purpose, we implement a discrete DRAM cache model which is flexible enough to cover a set of different DRAM cache designs.
This model relies on a separate DRAM cache manager which interacts with a near/fast/local memory controller (for a DRAM cache) and a far/slow/remote memory controller (for the backing store). This model does not require DRAM cache and the backing store to share a data bus.

A prior work proposed a unified DRAM cache controller to model the hardware of Intel's Cascade Lake~\cite{babaie2022cycle}.
It implements only one DRAM cache architecture where a DDR4 and NVM memory devices as the cache and the main memory are controlled by the same controller and share the data bus. This model is not flexible and generic to simulate different DRAM cache designs.
Our model in this work is able to cover this case. Since the discrete DRAM cache model is more flexible and can model future systems with DRAM cache designs, we will focus on that model in this paper.


\subsection{Background on \gem{}'s memory subsystem}

The \gem{} simulator is based on an event-driven full-system simulation engine.
It supports models of many system components, including memory controllers, memory device models, CPUs, and others.
The original memory controller module added to \gem{} by Hansson et al.~\cite{hansson2014simulating} is a \emph{cycle level} memory controller model designed to enable fast and accurate memory system exploration. The memory controller in \gem{} was refactored in~\cite{gem5-workshop-presentation} where two components were defined to simulate any controller.
(1) The \textit{memory controller} receives commands from the CPU and enqueues them into appropriate queues and manages their scheduling to the memory device.
(2) The \textit{memory interface} deals with device-specific timings and operations and communicates with the memory controller.
Moreover, the most recent release of \gem{} further improved the flexibility and modularity of memory controller and interfaces~\cite{akram2022modeling} and added HBM2 controller and interfaces.
Most importantly, it provides support for HBM2 interface and memory controller, where each physical channel is consisted of two pseudo-channels with peak theoretical bandwidth of 32 GB/s per channel.
The DRAM cache model presented in this work can employ any of the memory controller and interface model of \gem{} without any modifications.
Like \gem{}'s current memory controller, our DRAM cache model's goal is for cycle-level simulation to enable micro-architectural exploration and flexibility, not cycle-by-cycle accuracy to a single design.

\subsection{DRAM Cache Model}

\begin{figure}
  \centering
  \includegraphics[width=0.7\linewidth]{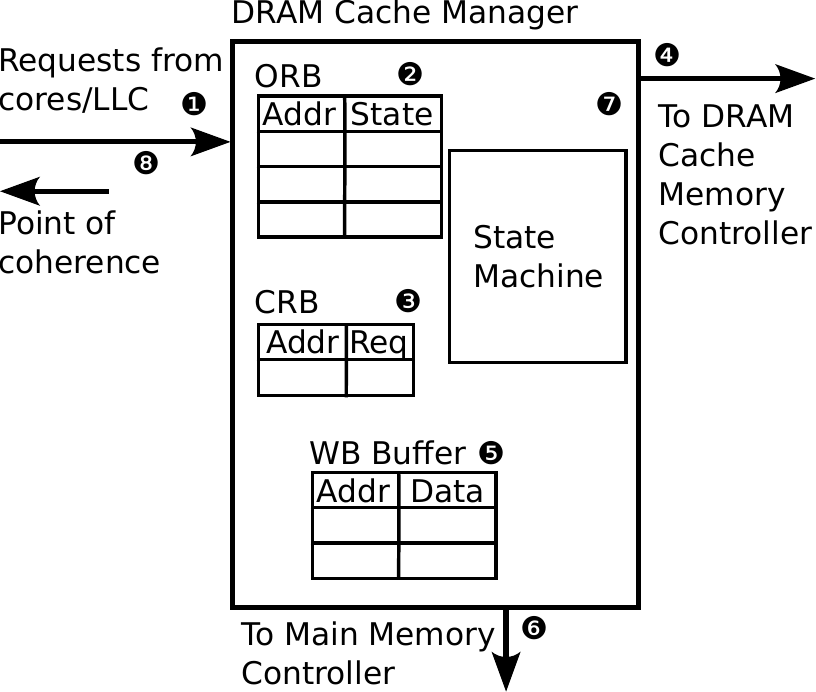}
  \vspace{-1ex}
  \caption{Overview of DRAM cache manager hardware.
  DRAM cache manager implements different DRAM caching policies and interacts with memory controllers of DRAM cache and its backing store.
  It includes three main buffers: (i) Outstanding Requests Buffer (ORB), (ii) Conflicting Requests Buffer (CRB), (iii) Write Back (WB) Buffer.
  }
  \label{fig:dcache}
  \vspace{-1.5em}
\end{figure}

Figure~\ref{fig:dcache} shows an overview of the DRAM cache model we implement in this work. DRAM cache manager receives all the incoming memory traffic (coming from the CPU, LLC, DMA).
This cache manager is not in the on-chip coherence domain and can be a drop-in replacement for the memory controller.

The DRAM cache manager is responsible for implementing different DRAM cache policies and interacts with two controllers.
In this work, near memory refers to the memory device which acts as a cache of the far memory (or the backing store).
The DRAM cache manager sends requests to and receives responses from the near and far memories. These requests include reads and writes
to the near and far memories and receiving a response for read requests and an acknowledgement for the write requests.
We allow the use of any memory controller in \gem{} as local and far memory controllers.
The \gem{} memory controllers take care of all the device-specific timing control and the DRAM cache manager allow us to isolate all DRAM cache specific controls.
This isolation makes the model modular and generic to implement different DRAM cache policies and architectures, as shown in the three case studies.

Our goal is to keep our DRAM cache model flexible enough to be able to simulate different DRAM cache designs.
Therefore, instead of modeling one specific micro-architecture with buffers for every device, we abstract away the hardware resources required to implement a DRAM cache controller by the use of two simple buffers: \textit{Outstanding Requests Buffer (ORB)} and \textit{Conflicting Requests Buffer (CRB)} shown in Figure~\ref{fig:dcache}.
All incoming memory requests reside in the \textit{ORB} unless a request conflicts with an already existing request in the \textit{ORB}.
Two requests are conflicting if they both map to the same location in the DRAM cache.
The conflicting request goes to the \textit{CRB} until the request it was conflicting with has been serviced and is taken out of the \textit{ORB}.
Each entry in these buffers contains other metadata in addition to the address of the request, as shown in Figure~\ref{fig:dcache}.
This metadata provides helpful information about the request, e.g., current state, and relative arrival time.
We also model a \textit{Write Back (WB) Buffer} for the DRAM cache dirty lines that are to be written back to the backing store.
What each entry in these buffers holds depend on the DRAM cache architecture and our model is flexible.





Figure~\ref{fig:dcache} also shows a state machine and a memory request goes through different steps while in the DRAM cache manager.
These steps depend on the DRAM cache design and the state machine contains the logic for implementing them.
Our model uses a tag and metadata storage in the DRAM cache manager to track the status of the DRAM cache locations.
This separate storage allows flexible implementation of different DRAM cache designs.
The model assures the timing implementation of the architecture as if this storage does not exist. Thus, the performance of each design is captured.

The steps in Figure~\ref{fig:dcache} represents an example of a DRAM cache design which is inspired by the real DRAM
cache implemented by Intel's Cascade Lake.
For the baseline architecture, we implement a DRAM cache which is direct-mapped (with caching granularity of 64 bytes), inserts on misses, and writes back the dirty cache lines upon evictions.
The tag and metadata are stored in ECC bits alongside the data.

Since \gem{} is an event driven simulator, the simulation model relies on scheduled events to transition between different states.
Below is an example of the state machine shown in Figure~\ref{fig:dcache}:

\paragraph*{Initializing a Request}
\textcircled{1} The CPU package (cores or LLC) sends a request to the main memory.
\textcircled{2} The DRAM cache manager (as the replacement for the memory controller) receives this request and places it in the ORB (if not conflicting with an existing request in the ORB). In case of writes, it sends an acknowledgement to the CPU/LLC.
\textcircled{3} If a conflict exists, the DRAM cache manager holds it in the CRB until the conflicting request leaves ORB.

\paragraph*{Request to the Local Memory}
\textcircled{4} Based on a scheduling policy (such as First-Come First-Serve) the DRAM cache manager sends a read request to the DRAM cache controller for tag check, and it receives
the response whenever it is ready. This response will provide the whole cache line that the demand access maps to in the DRAM cache, including data, tag and metadata.
If the tag matches, it is a hit, otherwise it is a miss on DRAM cache.
In case of read demand hit, the DRAM cache manager sends the response to the CPU/LLC.
In case of write demand hit or miss, the DRAM cache manager sends a write request with the data from demand access to the DRAM cache controller.
\textcircled{5} If any of the misses had a dirty flag set in the metadata, the DRAM cache manager inserts that cache line into the WB buffer.
Whenever there is bandwidth available or if the WB buffer becomes full, the DRAM cache manager sends these write backs as a write request to the main memory controller.

\paragraph*{Request to the Remote Memory}
\textcircled{6} In case of read demand miss, the DRAM cache manager sends a read request to the main memory controller to fetch the missing cache line from backing store.
Once the DRAM cache manager receives the response from main memory controller, it finishes the miss handling process.
First, \textcircled{7} it sends a write request to the DRAM cache controller to fill the missing cache line.
Second, \textcircled{8} it sends the response for the demand to CPU/LLC.

\section{Methodology}
\label{method}


\begin{figure}
  \centering
  \subfloat[Target AMD EPYC-like system.]{
    \centering
  \includegraphics[width=0.44\linewidth]{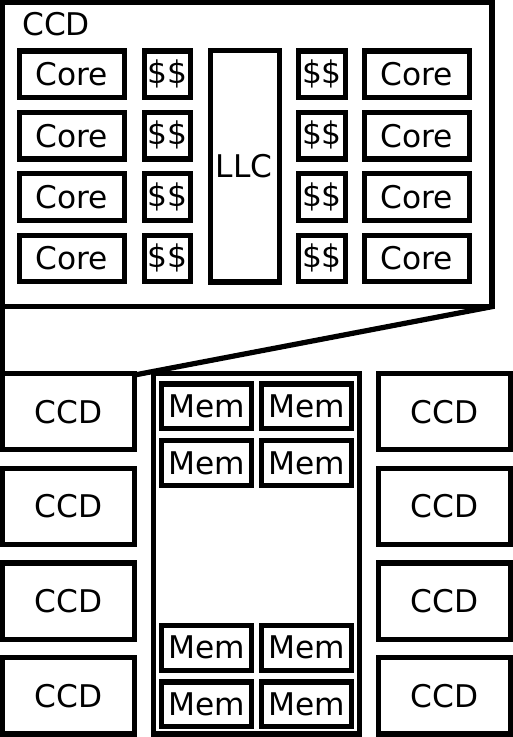}
  \label{fig:whole-system}
  }
  \hfill
  \subfloat[A $\nicefrac{1}{8}$ system used in simulation. The system used in each case study is shown.]{
    \centering
  \includegraphics[width=0.48\linewidth]{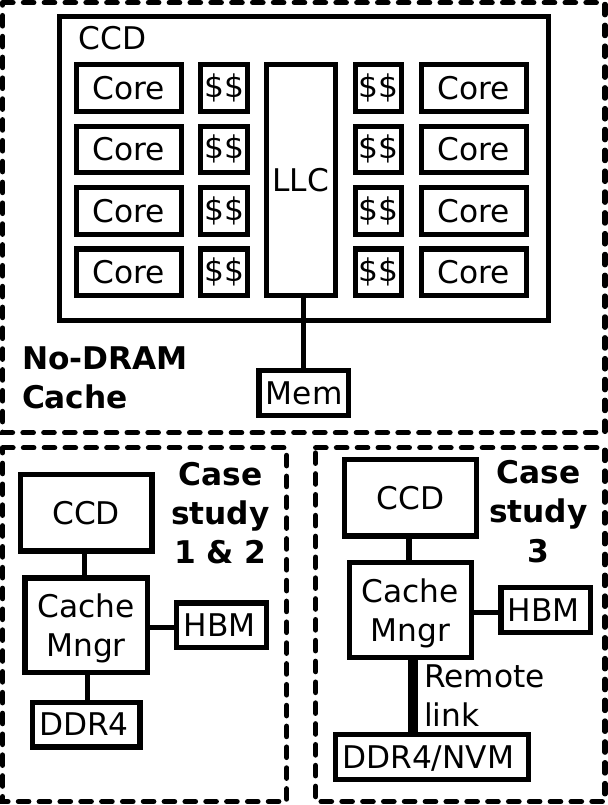}
  \label{fig:case-studies}
  }
  \caption{(a) Target system: AMD EPYC-like system, which contains eight cores per CCD. CCDs are connected to the eight main memory channels via an I/O die.
  The memory system contains an HBM stack which can act as a cache to the 8-channel DDR4-based main memory.
  We model $\nicefrac{1}{8}$ as shown on the right.}
  \label{fig:system}
  \vspace{-1.5em}
\end{figure}

\subsection{Modeled System}
Figure~\ref{fig:whole-system} shows the target system we simulated for our experiments.
We consider $\nicefrac{1}{8}$ of a single socket of an AMD EPYC-like system~\cite{naffziger2021pioneering} which consists of a single
core complex die (CCD) and a single channel of main memory, as shown in Figure~\ref{fig:case-studies} on the top. The modeled CCD contains eight cores that have private L1 caches and share the last level cache among the cores.
The main reason for our decision to model $\nicefrac{1}{8}$ of the system is that with the current core and on-chip cache models of \gem{}, it will take a long simulation time to model the entire system in a simulated environment.

To incorporate our DRAM cache model with this system, we replace the memory controller with a DRAM cache manager,
as shown in Figure~\ref{fig:case-studies} on the bottom. For the baseline, the cache manager uses a single channel of HBM2
(which consists of two pseudo-channels and acts as a DRAM cache), and a single channel of DDR4 (which serves as the main memory in the system).
Note that in these figures the memory interfaces and their controllers are shown in the same box. Table \ref{tab:baseConfig} summarizes the baseline system configuration.


\begin{table}[!h]
  \begin{center}
  \caption{\label{tab:baseConfig}Baseline System Configuration}
  \begin{tabular}{ |p{3.5cm}||p{2.5cm}|}
    \hline
    \multicolumn{2}{|c|}{Processors} \\
    \hline
    Number of cores & 8\\
    Frequency & 5 GHz\\
    \hline
    \multicolumn{2}{|c|}{On-chip Caches} \\
    \hline
    Private L1 Instruction & 32 KB\\
    Private L1 Data & 512 KB\\
    Shared L2 & 8 MB\\
    \hline
    \multicolumn{2}{|c|}{DRAM Cache Manager} \\
    \hline
    ORB & 128 entries\\
    CRB & 32 entries\\
    WB Buffer & 64 entries\\
    Frontend/Backend Latencies& 20 ns round-trip\\
    \hline
    \multicolumn{2}{|c|}{DRAM Cache (HBM2)} \\
    \hline
    Capacity & 128 MB\\
    Theoretical Peak Bandwidth & 32 GB/s\\
    Read/Write Buffer & 64 entries each\\
    \hline
    \multicolumn{2}{|c|}{Main Memory (DDR4/NVM)} \\
    \hline
    Capacity & 3 GB\\
    Theoretical Peak Bandwidth & 19.2 GB/s\\
    Read/Write Buffer & 64 entries each\\
    \hline
  \end{tabular}
  \end{center}
  \end{table}

\subsection{Workloads used}
We evaluated DRAM caches using a subset of multithreaded workloads in NPB~\cite{bailey1991parallel} and GAPBS~\cite{beamer2015gap} that assess high-performance systems.
We use the C class of the NPB workloads and a synthetic graph as an input (size of $2^{22}$ vertices) for the GAPBS workloads.
These workloads' working-set sizes varies between few hundreds MB to 1 GB. Thus, we set the size of the DRAM cache and the main memory to 128 MB and 3 GB, respectively, so the DRAM cache is smaller than the workloads' memory footprints.

\subsection{Simulation methodology for benchmarks}
\begin{figure}
    \centering
    \frame{\includegraphics[width=0.7\linewidth]{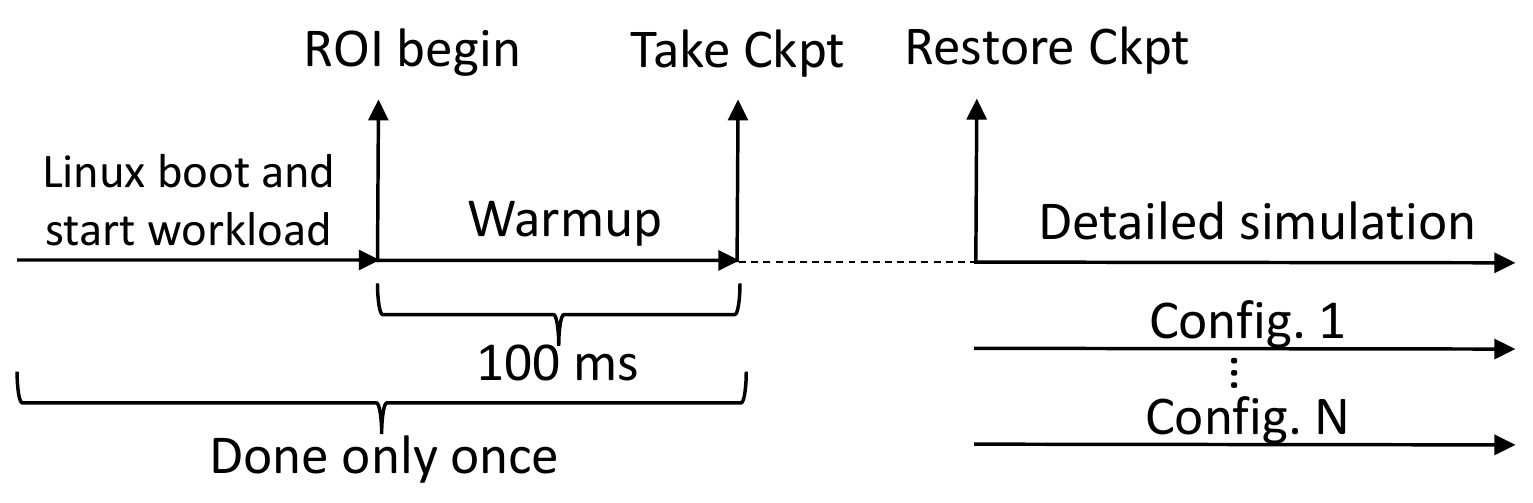}}
    \caption{Summary of the simulation methodology. Ckpt: checkpoint, ROI: region of interest. We use a warm-up period of 100ms and detailed simulation time of 1s.}
    \label{fig:method}
    \vspace{-1.5em}
\end{figure}

Figure~\ref{fig:method} shows our methodology for the experiments we ran in this paper.
First, Linux kernel boots on the target system in \gem{} and the execution of the program starts and continues until the start of the region of interest (ROI) of the workload. The benchmarks are marked with ROI begin markers using \gem{} pseudo instruction support.
Beginning at the ROI, we simulate the workload for 100ms to warm up the system including the DRAM cache. At the end of 100ms, we take a checkpoint. This process is done once per workload. Later, we restore from the checkpoint to run all our simulations with different DRAM cache configurations.
Using a checkpoint ensures that all of our runs have the same starting point with the same system state for a fair comparison across different tested configurations.
We simulate the restored checkpoint for either one second of simulation time or reaching the end of ROI, whichever comes first.
Our results show an average cold-miss ratio of 3.5\% for DRAM cache, during the restore simulation.

\begin{figure}
  \centering
  \includegraphics[width=0.8\linewidth]{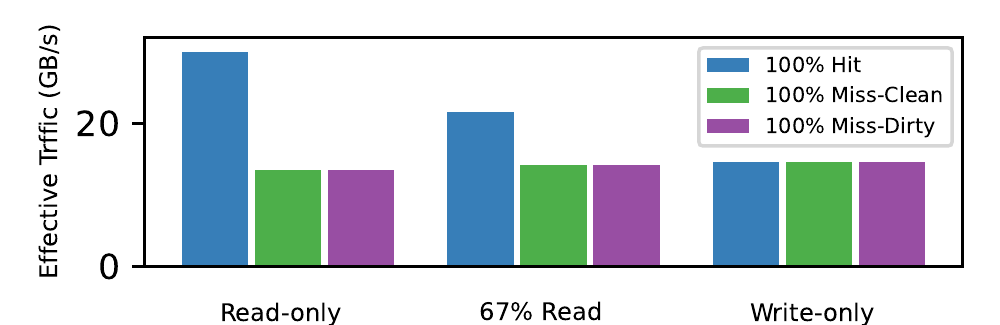}

  \caption{Effective traffic at the LLC. Synthetic traffic was injected to the DRAM cache manager.
  The traffic was controlled for percentage of read/writes, miss ratio, and clean or dirty line eviction ratio.}
  \label{fig:trafGen}
  \vspace{-1.5em}
\end{figure}

Figure~\ref{fig:trafGen} demonstrates a validation of our DRAM cache model. We used a traffic generator to replace LLC/CPU side in Figure~\ref{fig:dcache}. This traffic generator creates synthetic memory traffic configurable for read/write percentage, random/linear pattern, etc. We used a single physical channel of HBM2 (32 GB/s peak bandwidth) as the DRAM cache and a single channel of DDR4 (19.2 GB/s peak bandwidth) as the main memory. We controlled the miss ratio,
percentage of read requests, and the ratio of dirty or clean cache lines in case of misses. We ran the tests for 10 ms, enough to reach the saturated bandwidth in all cases.
For read-only (RO) traffic with a 100\% hit ratio, the DRAM cache should perform the same as the main memory, which is observed in Figure~\ref{fig:trafGen}. DRAM cache shows effective traffic of 29.94 GB/s (similar to \gem{} HBM2 memory controller effective bandwidth).
As we add more writes to the accesses and test with 67\% read (33\% write) and 100\% write patterns, the effective bandwidth drops as write hits require two accesses to the DRAM cache.
The effective bandwidth drops further for the 100\% miss ratio due to more extra accesses each demand request needs to make to the DRAM cache and the backing store.
For the Miss-Dirty case, there is a write-back to the main memory. However,
the main memory controller has enough bandwidth to handle these writes while the DRAM cache
pipeline is saturated.

\section{Case Study 1: Performance of Baseline Cache}
\label{cs1}


In this case study, we will investigate the performance and different pathologies memory requests go through, in the baseline DRAM cache design.
Specifically, we ask \emph{how a DRAM cache system's performance compares to those without a DRAM cache.} For this purpose, we consider the simulated system described in Section~\ref{method} in two different configurations: 1) HBM2 as a DRAM cache with DDR4 as main memory, and 2) no DRAM cache and DDR4 as main memory.
We run all workloads in these two configurations.

Typically, we would expect an HBM2-based DRAM cache to perform better, specifically in comparison to DDR4 main memory, because of its higher bandwidth. This case study considers the scenario where there is no additional latency, and the main memory is close to the local memory (e.g., connected to same package).
The similar latency of HBM2 and DDR4 might result in low-performance improvements with the HBM2 DRAM cache. In Section~\ref{sec:cs3}, we will evaluate a more realistic scenario of a high latency to main memory (e.g., disaggregated systems).
We expect that the DRAM caches perform similarly to a DRAM main memory if the working set of the workload fits in the cache (or has a high DRAM cache hit rate). Therefore, rather than focus on the obvious case, we evaluate the performance of the DRAM cache-based system when the workload does not fit in the DRAM cache. We assess the performance of selected workloads on the target system discussed in Section~\ref{method} to accomplish the previously mentioned goal.
We compare the amount of work each configuration has done during its execution to evaluate its performance.

\begin{figure}

    \subfloat[Performance]{%
      \includegraphics[clip,width=\linewidth]{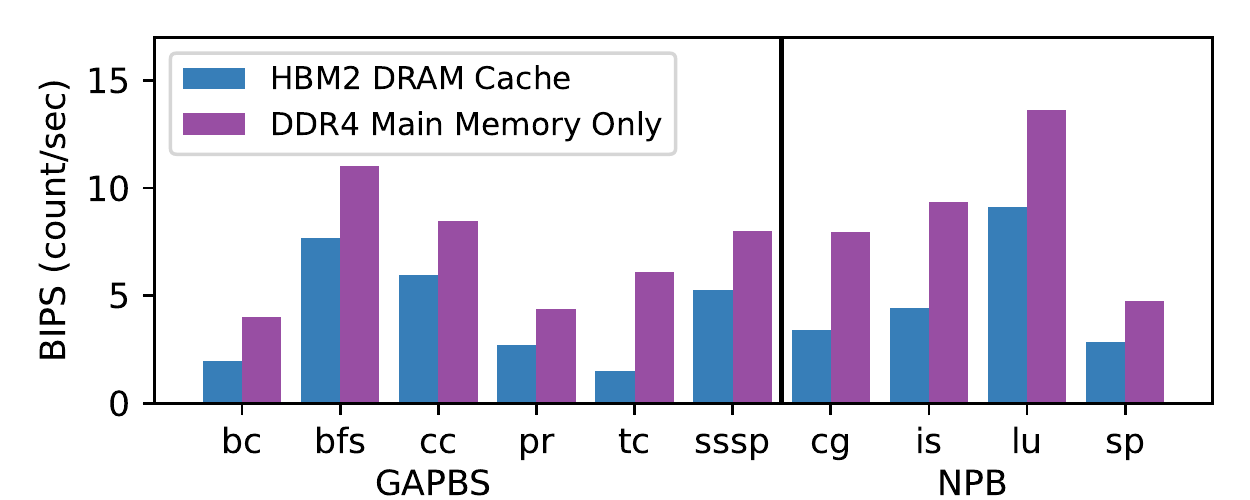}%
      \vspace{-3ex}
      \label{fig:cs1Bips}
    } \\
    \subfloat[Speedup]{%
      \includegraphics[clip,width=\linewidth]{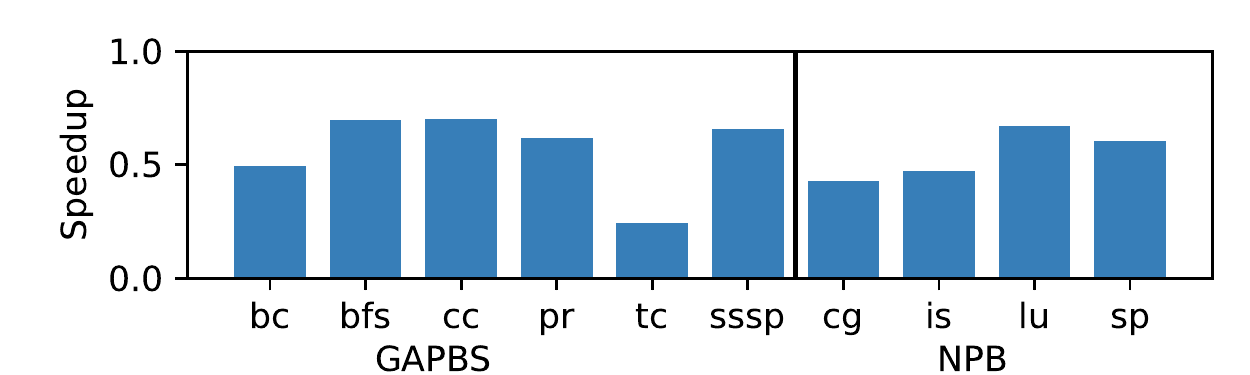}%
      \label{fig:cs1Sup}
    }
    \caption{Performance comparison of the baseline DRAM cache to the same system without the DRAM cache, based on billion instructions per second (BIPS). DRAM cache based system always performs less than the system without DRAM cache.
    \string*$bt$ is excluded from this analysis, as it started a different phase of program execution.}
    \label{fig:cs1BipsSu}
    \vspace{-1.5em}

\end{figure}

%

\begin{figure}
    \centering
    \includegraphics[width=\linewidth]{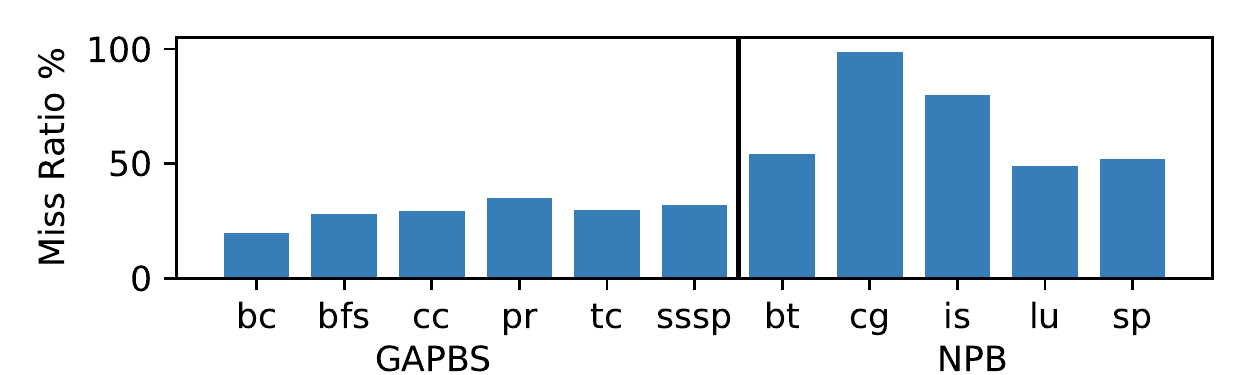}
    \vspace{-1ex}
    \caption{Miss ratio of different workloads on DRAM cache. Miss ratios range
    from 20\% to 100\% approximately.}
    \label{fig:cs1MissRatio}
    \vspace{-1.5em}
\end{figure}

\subsection*{Results and Discussions}
Figure~\ref{fig:cs1Bips} shows the performance for NPB and GAPBS workloads in the two configurations we described above. We use a metric of billion instructions per second (BIPS) to represent performance.
Figure~\ref{fig:cs1Sup} shows the speed-up achieved with a DRAM cache-based system compared to a system without a DRAM cache (and DDR4 main memory only).
We expect DRAM caches to perform better than a far memory alone.
However, as Figure~\ref{fig:cs1Sup} shows, DRAM cache configuration performs worse than just a DDR4 main memory for all these tests (speed-up is consistently below 1).
We observe that the DRAM caches start to perform poorly as the DRAM cache miss rate is above 20\% (in the workloads we ran). Figure~\ref{fig:cs1MissRatio} shows the miss ratio of DRAM cache per workload.
We define the miss ratio as the total number of miss accesses to the DRAM cache divided by the sum of total miss and hit accesses.
As shown in Figure~\ref{fig:cs1MissRatio}, GAPBS workloads have miss ratios ranging from 20\% to 50\%, and NPB has miss ratios mostly around 50\%. These miss ratios are high but not unreasonable, and we can expect real-life workloads on high-performance systems to show such miss ratios~\cite{hildebrand2021case}.
As described in Section~\ref{design}, miss handling in current hardware-managed DRAM caches can lead to multiple accesses to the cache (local) and backing store (far) interfaces.

This mostly serialized process cannot fully utilize the available bandwidth at the local and far interfaces and
affects to the overall system performance by increasing the latency.
~The interference between these extra accesses and the actual demand access leads to performance degradation, and the previous
DRAM cache studies were not able to capture it. Note that we made sure that the DRAM cache had been warmed-up, so cold misses
are not contributing to the performance observed from the system.
Note: $bt$ in NPB is an outlier in this evaluation because our analysis shows it started another phase of execution. Thus, we exclude it from this analysis.

\begin{figure}
    \centering
    \includegraphics[width=\linewidth]{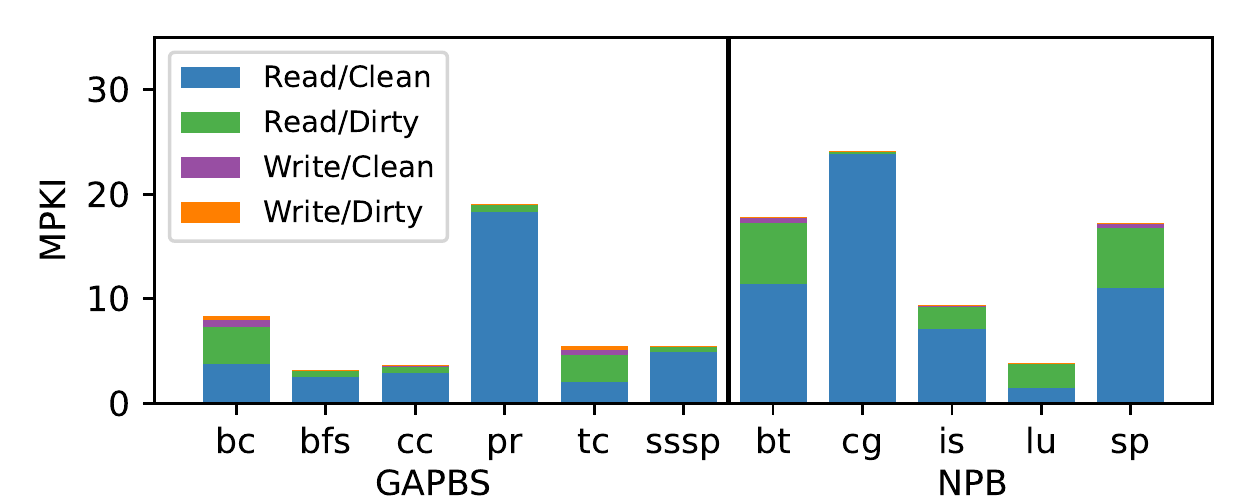}
    \vspace{-1ex}
    \caption{DRAM cache misses per thousand instructions for GAPBS and NPB.
    The figure also shows the distribution of different miss cases out of the total number of miss accesses.}
    \label{fig:cs1Mpki}
    \vspace{-1.5em}
\end{figure}

\begin{figure}
    \centering
    \includegraphics[width=\linewidth]{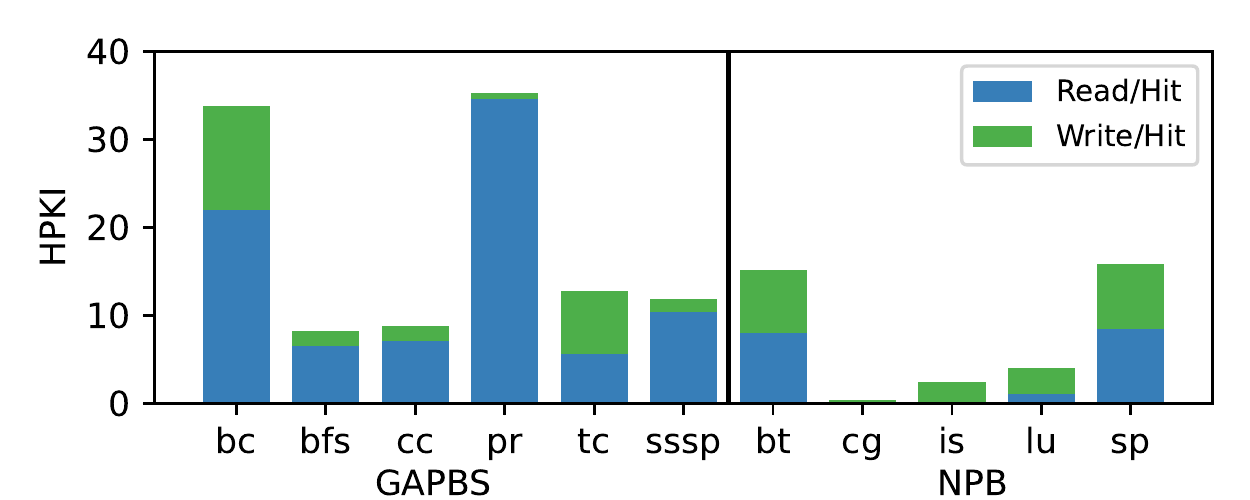}
    \vspace{-1ex}
    \caption{DRAM cache hits per thousand instructions for GAPBS and NPB.
    The figure also shows the distribution of read and write hits out of the total number of hit accesses.}
    \label{fig:cs1Hpki}
    \vspace{-1.5em}
\end{figure}

In addition to the total number of DRAM cache misses, we must look at the types of misses to fully explain the DRAM cache slowdown.
Figure~\ref{fig:cs1Mpki} shows the misses per thousand instruction (MPKI), and the breakdown of the misses into different types in terms of
read or write request and whether the cache location that the request mapped to was clean or dirty.
The type of DRAM cache miss determines how many extra accesses or operations need to be performed for the demand access under consideration.
For example, $tc$ and $bc$ in Figure~\ref{fig:cs1Sup} show the highest slowdown among the GAPBS workloads as they have the highest fraction of dirty misses (Figure~\ref{fig:cs1Mpki}). Similar behavior is true for $sp$ from NPB benchmarks.
$pr$ from GAPBS and $cg$ from NPB do not have a high fraction of dirty cache misses but still show significant DRAM cache slowdown as the total number of misses demonstrated by these benchmarks is relatively high.
Figure~\ref{fig:cs1Hpki} shows hits per thousand instructions (HPKI) and the read/write distribution of DRAM cache hits.
Since write hits have one more access to the DRAM than read hits, even write hits can cause performance degradation. For example, $tc$ and $bc$ from GAPBS have a high fraction of write hits contributing to their large slowdowns.

\begin{figure}
    \centering
    \includegraphics[width=\linewidth]{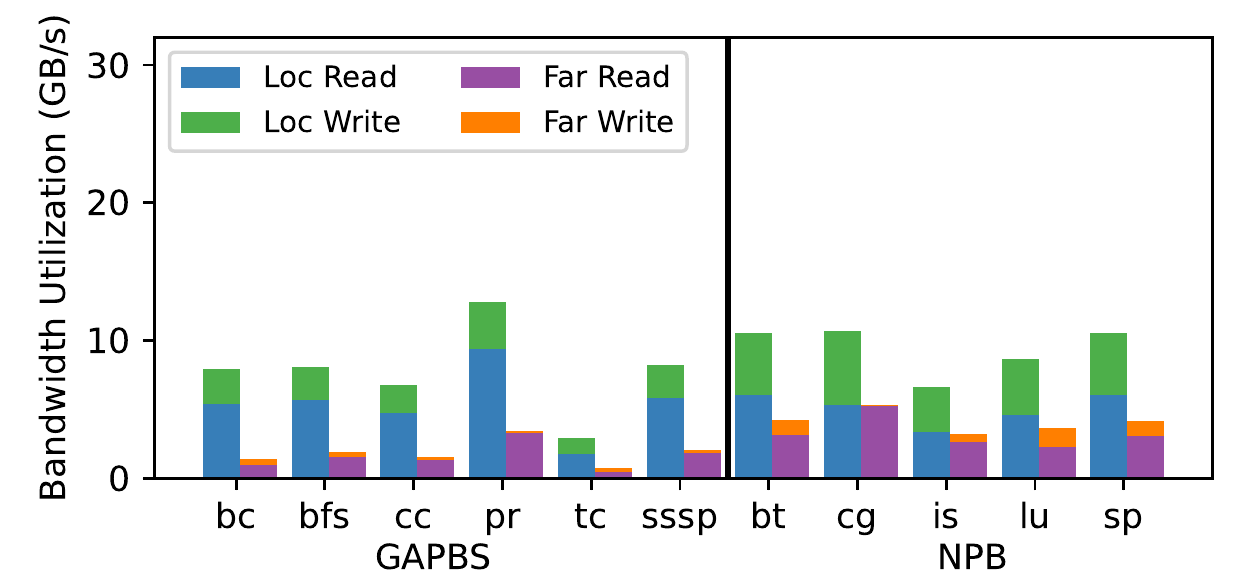}
    \vspace{-1ex}
    \caption{Bandwidth utilization of different workloads on DRAM cache (local) and backing store (far) interfaces. Peak bandwidth is 32 GB/s and 19.2 GB/s for local and far memory. Figure shows that the workloads significantly under-utilize the available bandwidth due to access amplification.}
    \label{fig:cs1BwUtil}
    \vspace{-1.5em}
\end{figure}

Figure~\ref{fig:cs1BwUtil} shows the bandwidth utilization per workload for each memory interface (local and far memory). The HBM2 DRAM cache
interface and the DDR4 backing store interface of \gem{} used in this study have a theoretical peak bandwidth of 32 GB/s and 19.2 GB/s, respectively.
Figure~\ref{fig:cs1BwUtil} shows that for all the cases, a significant amount of bandwidth remains unused for both memory interfaces. The reason for low utilization is the latency cost because of extra memory accesses needed in case of a DRAM cache miss (for both read and write), or a DRAM cache write hit.

\textbf{Takeaway:} the current DRAM cache designs perform poorly even when the DRAM cache miss ratio is as low as 20\%. In case of DRAM cache miss, the extra memory accesses required for single demand access lead to increased access latency and under-utilization of available bandwidth to the memory devices.
This behavior shows that we need to optimize the DRAM cache design for both miss and write-hit handling to improve the performance.

\section{Case Study 2: Cache Architecture Exploration}
\label{sec:cs2}

In previous section we showed the performance degradation of the baseline system with DRAM cache (Figure \ref{fig:case-studies} bottom left) compared to the same system without the DRAM cache (Figure \ref{fig:case-studies} top). In this study,
we investigate optimized DRAM cache designs to mitigate the overheads shown in Section \ref{cs1}.
The baseline DRAM cache protocol of our model which is used in Section \ref{cs1} is inspired by the real hardware
of Intel's Cascade Lake. However, our model is capable of implementing different DRAM cache
designs. To show this capability, in this section we show two different cache architecture optimizations
on top of the baseline model.


As explained in the Section \ref{design}, the baseline DRAM cache architecture stores the tag and metadata in ECC bits in the cache line, along with the data.
Thus, to check whether a memory request is hit or miss on the DRAM cache, it reads the entire cache line from DRAM, while
only the tag and metadata is needed for tag comparison. If the memory request is miss clean (for both read and write requests) or write hit on DRAM cache,
reading the data becomes a waste of bandwidth. We elaborate the details of these cases as follows:

\paragraph {Write Hits}
Figure \ref{fig:cs1Hpki} shows write hit requests consist a noticeable
portion of total hit accesses. For these accesses in the baseline design, cache manager reads the cache line from DRAM cache for tag comparison.
Once the cache manager receives the cache line and the tag in the request address and the tag in the cache line match, it sends a write request with the data from demand access to the cache line address.
In this process, the data in the cache line fetched for tag check is never used and is overwritten, eventually. Figure \ref{fig:cs1Hpki} shows 
write hit accesses consist a large part of all hit accesses.

\paragraph {Read or Write Miss-Cleans}
As we described above for write hits, the DRAM cache manager reads the entire cache line that the request maps to for tag check.
Once the tag comparison in the DRAM cache manager fails to match, if the request is a write, the cache manager sends a write request with the data from the demand access to the DRAM cache line address.
If the request is read, the DRAM cache manager sends a read request to the backing store to get the missing cache line from the main memory.
In these processes, the data in the cache line which was fetched for tag check is never used. 
Figure \ref{fig:cs1Mpki} shows that significant portions of the total memory requests that miss on DRAM cache consist of miss accesses to clean cache lines.

Given these two scenarios, where there is a bandwidth waste by the baseline DRAM cache, we investigate two different designs for the DRAM cache state machine.
First, we implement the optimization introduced by the BEAR cache for some write accesses~\cite{chou2015bear}.
BEAR cache tries to avoid reading the line from DRAM cache for tag check, for the write hit accesses to the DRAM cache.
BEAR determines the write hit accesses using the metadata stored in the last level cache.
We apply this optimization to the baseline architecture of our model and name this case \emph{\bear{}}.

Second, we change the baseline DRAM cache design in a way that avoids the read access to DRAM cache for tag
check, not only for write hit demands, but also if the demand access (either read or write) will miss on DRAM cache and the cache line is
clean. We call this case \emph{\oracle{}}. We assume \oracle{} has a zero latency SRAM storage in the cache manager to hold all the tag and metadata
it needs to determine if the demand access will hit or miss to a clean or dirty cache line. Table \ref{tab:accAmp} summarizes the optimizations of \bear{} and \oracle{} compared to the baseline in terms of
reducing extra accesses from pathology of DRAM cache design in all possible memory request cases.

To compare the performance of these three different designs (baseline, \bear{}, and \oracle{}), we run the GAPBS and NPB through full-system simulation
as explained in Section \ref{method}. In all three cases, the cache manager uses HBM2 DRAM cache and DDR4 main memory.

\begin{table}[!h]
    \begin{center}
    \caption{\label{tab:accAmp} Comparison of number of extra accesses needed for a given memory request in
    Baseline, \bear{}, and \oracle{} designs, written in black, blue, and red, respectively.}
    \resizebox{\linewidth}{!}{
    \begin{tabular}[width=\linewidth]{| c | c | c | c | c | c | c | c | c |}
      \hline
      Access & \multicolumn{4}{c|}{Read} & \multicolumn{4}{c|}{Write}\\
      \hline
      Hit/Miss & \multicolumn{2}{c|}{Hit} & \multicolumn{2}{c|}{Miss} & \multicolumn{2}{c|}{Hit} & \multicolumn{2}{c|}{Miss}\\
      \hline
      Dirty/Clean & \multicolumn{1}{c|}{Dirty} & \multicolumn{1}{c|}{Clean} & \multicolumn{1}{c|}{Dirty} & \multicolumn{1}{c|}{Clean} & \multicolumn{1}{c|}{Dirty} & \multicolumn{1}{c|}{Clean} & \multicolumn{1}{c|}{Dirty} & \multicolumn{1}{c|}{Clean}\\
      \hline
      Local Read &
       \ding{51} \textcolor{blue}{\ding{51}} \textcolor{red}{\ding{51}} &
       \ding{51} \textcolor{blue}{\ding{51}} \textcolor{red}{\ding{51}} &
       \ding{51} \textcolor{blue}{\ding{51}} \textcolor{red}{\ding{51}} &
       \ding{51} \textcolor{blue}{\ding{51}} \textcolor{red}{\ding{55}} &
       \ding{51} \textcolor{blue}{\ding{55}} \textcolor{red}{\ding{55}} &
       \ding{51} \textcolor{blue}{\ding{55}} \textcolor{red}{\ding{55}} &
       \ding{51} \textcolor{blue}{\ding{51}} \textcolor{red}{\ding{51}} &
       \ding{51} \textcolor{blue}{\ding{51}} \textcolor{red}{\ding{55}} \\
      \hline
      Far Read &
      \ding{55} \textcolor{blue}{\ding{55}} \textcolor{red}{\ding{55}} &
      \ding{55} \textcolor{blue}{\ding{55}} \textcolor{red}{\ding{55}} &
       \ding{51} \textcolor{blue}{\ding{51}} \textcolor{red}{\ding{51}} &
       \ding{51} \textcolor{blue}{\ding{51}} \textcolor{red}{\ding{51}} &
       \ding{55} \textcolor{blue}{\ding{55}} \textcolor{red}{\ding{55}} &
       \ding{55} \textcolor{blue}{\ding{55}} \textcolor{red}{\ding{55}} &
       \ding{55} \textcolor{blue}{\ding{55}} \textcolor{red}{\ding{55}} &
       \ding{55} \textcolor{blue}{\ding{55}} \textcolor{red}{\ding{55}} \\
      \hline
      Local Write &
      \ding{55} \textcolor{blue}{\ding{55}} \textcolor{red}{\ding{55}} &
      \ding{55} \textcolor{blue}{\ding{55}} \textcolor{red}{\ding{55}} &
       \ding{51} \textcolor{blue}{\ding{51}} \textcolor{red}{\ding{51}} &
       \ding{51} \textcolor{blue}{\ding{51}} \textcolor{red}{\ding{51}} &
       \ding{51} \textcolor{blue}{\ding{51}} \textcolor{red}{\ding{51}} &
       \ding{51} \textcolor{blue}{\ding{51}} \textcolor{red}{\ding{51}} &
       \ding{51} \textcolor{blue}{\ding{51}} \textcolor{red}{\ding{51}} &
       \ding{51} \textcolor{blue}{\ding{51}} \textcolor{red}{\ding{51}} \\
      \hline
      Far Write &
      \ding{55} \textcolor{blue}{\ding{55}} \textcolor{red}{\ding{55}} &
      \ding{55} \textcolor{blue}{\ding{55}} \textcolor{red}{\ding{55}} &
       \ding{51} \textcolor{blue}{\ding{51}} \textcolor{red}{\ding{51}} &
       \ding{55} \textcolor{blue}{\ding{55}} \textcolor{red}{\ding{55}} &
       \ding{55} \textcolor{blue}{\ding{55}} \textcolor{red}{\ding{55}} &
       \ding{55} \textcolor{blue}{\ding{55}} \textcolor{red}{\ding{55}} &
       \ding{51} \textcolor{blue}{\ding{51}} \textcolor{red}{\ding{51}} &
       \ding{55} \textcolor{blue}{\ding{55}} \textcolor{red}{\ding{55}} \\
      \hline
      \hline
      Tot. Baseline & 1 & 1 & 4 & 3 & 2 & 2 & 3 & 2 \\
      \hline
      Tot. \bear{} & \textcolor{blue}{1} & \textcolor{blue}{1} & \textcolor{blue}{4} & \textcolor{blue}{3} & \textcolor{blue}{1} & \textcolor{blue}{1} & \textcolor{blue}{3} & \textcolor{blue}{2} \\
      \hline
      Tot. \oracle{} & \textcolor{red}{1} & \textcolor{red}{1} & \textcolor{red}{4} & \textcolor{red}{2} & \textcolor{red}{1} & \textcolor{red}{1} & \textcolor{red}{3} & \textcolor{red}{1} \\
      \hline
    \end{tabular}
    }
    \end{center}
\end{table}

\subsection*{Results and Discussions}

\begin{figure}
    \centering
    \includegraphics[width=\linewidth]{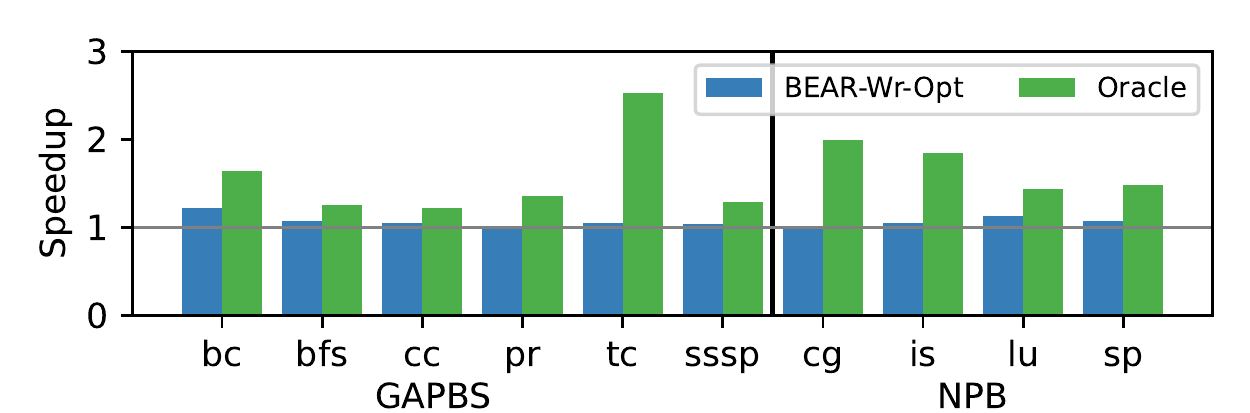}
    \caption{Speedup of \bear{} and \oracle{} compared to the baseline DRAM cache design in GAPBS and NPB.
    \string*$bt$ is excluded from \oracle{} analysis, as it started a different phase of program execution.}
    \label{fig:cs2SuAll}
    \vspace{-1.5em}
\end{figure}

Figure \ref{fig:cs2SuAll} shows the performance speedup of \oracle{} and \bear{} compared to the baseline.
\oracle{} and \bear{} outperform the baseline architecture. The speedup gain for
\oracle{} is higher compared to \bear{}, because the optimization of \oracle{} is more extensive than \bear{}, and
it optimizes more demand access pathologies in DRAM cache protocol, compared to \bear{}.

\subsection{GAPBS in \bear{}}

$bc$ has a significant number of write hits in its hits per thousand instructions or HPKI (Figure \ref{fig:cs1Hpki}).
$bc$ also has the lowest miss ratio of all the workloads in GAPBS (Figure \ref{fig:cs1MissRatio}).
Removing the DRAM read for tag check from write hit handling path (what \bear{} does) reduces the latency of servicing misses
and increases system throughput.
Figure \ref{fig:cs2SuAll} shows $bc$ has the highest speedup amongst all workloads in GAPBS for \bear{}.

$pr$ has the lowest amount of speedup for \bear{} because it has the least amount of write hits in its HPKI
(Figure \ref{fig:cs1Hpki}) and it has the highest miss ratio (Figure \ref{fig:cs1MissRatio}).
Optimizing a non-significant number of write hits in this workload does not reduce the latency of misses.
As a result, the speedup is hardly above one.

Another interesting case is $tc$ which has a significant number of write hits in its HPKI (Figure \ref{fig:cs1Hpki});
however, it does not benefit from \bear{}.
The reason is $tc$ has significantly higher ratio of miss dirty lines, than the rest of workloads in GAPBS.
All the dirty evicted lines must be written back to the main memory. DRAM cache manager holds them in Write Back buffer
(WB buffer in Figure \ref{fig:dcache}). Once the WB buffer is full, the cache manager prioritizes the writes in
the WB buffer over the requests in outstanding requests buffer (ORB in Figure \ref{fig:dcache}).
In this way, the cache manager makes entry available in the WB buffer for incoming DRAM cache line
evictions that are dirty. This prioritization can increase the latency of the requests in ORB.
Even though the write hit optimization in \bear{} can reduce the latency of misses, for a write intensive application
with relatively high miss ratio like $tc$ is not beneficial.
The rest of the workloads’ speedups are proportional to their write hits and miss ratios.

\subsection{NPB in \bear{}}

Figure \ref{fig:cs2SuAll} shows $lu$ has the highest speedup in NPB for \bear{}.
$lu$ has a large number of write hits in its HPKIs (Figure \ref{fig:cs1Hpki})
and the lowest miss ratios compared to the rest of workloads in NPB (Figure \ref{fig:cs1MissRatio}).
As we explained above, removing the DRAM read for tag check from write hit handling path in \bear{},
reduces the latency of servicing misses and increases system throughput.

$cg$ dose not benefit from \bear{} since it has close to 100\% miss ratio (Figure \ref{fig:cs1MissRatio})
and very limited number of write hits (Figure \ref{fig:cs1Hpki}).

\subsection{GAPBS in \oracle{}}

Figure \ref{fig:cs2SuAll} shows $tc$ and $bc$ have the highest speedup amongst all workloads for \oracle{}.
$tc$ and $bc$ have highest portion of write hits in their HPKIs (Figure \ref{fig:cs1Hpki}), also,
lower miss ratios compared to the other workloads in the suite (Figure \ref{fig:cs1MissRatio}).
Moreover, they have the highest portion of write miss cleans in their Misses per Thousand Instructions or MPKI
(Figure \ref{fig:cs1Mpki}) which can have an out of size impact on the performance improvement.
The speedups of the other workloads in GAPBS is proportional to the number of read miss cleans accesses (Figure \ref{fig:cs1Mpki})
that \oracle{} optimizes.

\subsection{NPB in \oracle{}}

The MPKIs of $cg$, $is$, and $sp$ workloads are dominated by read miss cleans (Figure \ref{fig:cs1Mpki})
and they have the highest miss ratios (Figure \ref{fig:cs1MissRatio}).
Removing the DRAM read for tag check from read miss cleans handling path (what \oracle{} does)
reduces their latencies and increases system throughput. Thus, $cg$, $is$, and $sp$ have the highest
speedups amongst all workloads as Figure \ref{fig:cs2SuAll}. $lu$ has the lowest speedup because its MPKI is dominated by read miss dirties, not optimized by \oracle{}.

$bt$ in NPB is an outlier in this evaluation because our analysis shows it started another phase of execution. Thus, we exclude it from analysis of \bear{} and \oracle{}.

\begin{figure}
  \centering
  \includegraphics[width=\linewidth]{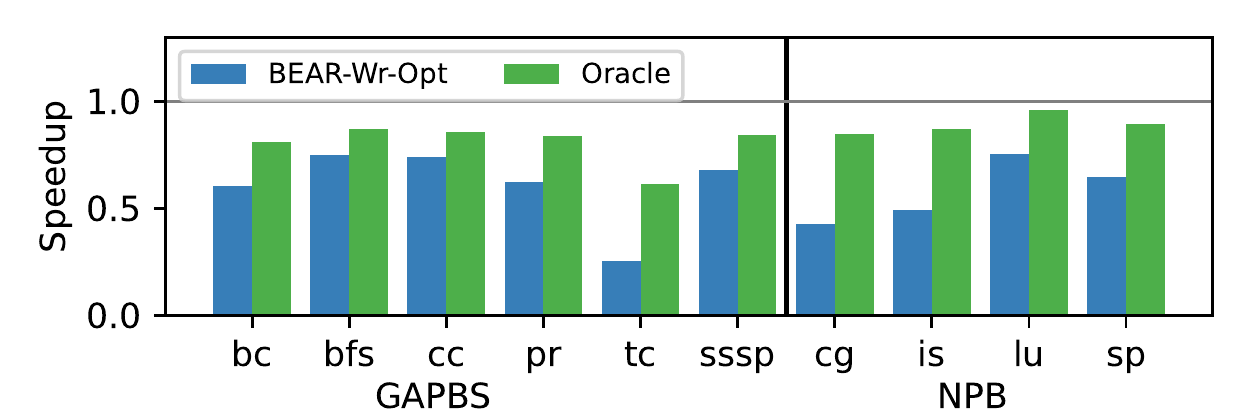}
  \caption{Speedup of \bear{} and \oracle{} compared to the system without the DRAM cache in GAPBS and NPB.
  \string*$bt$ is excluded from \oracle{} analysis, as it started a different phase of program execution.}
  \label{fig:cs2SuAllMm}
  \vspace{-1.5em}
\end{figure}

Finally, we show the speedup of \bear{} and \oracle{} compared to the same system
without the DRAM cache (with a DDR4 main memory only) in Figure~\ref{fig:cs2SuAllMm}. We observe that both optimized DRAM cache designs we investigate
in this case study still perform less than a system without the DRAM cache. \oracle{} optimizes more pathologies
in the baseline DRAM cache design compared to \bear{}; thus, it has higher speedup than \bear{}. Even though
\oracle{} uses a zero-latency SRAM tag and metadata storage in its implementation, it still needs further optimization
to outperform a system without the DRAM cache.

\textbf{Takeaway:} 
In this case study we showed if a cache design can provide a mechanism to determine the hit/miss and clean/dirty for a given request without accessing the DRAM,
it can provide performance improvement compared to the baseline DRAM cache design. However, these optimizations 
are not enough for the DRAM cache system to outperform the same system without the DRAM cache.


\section{Case Study 3: Impact of Link Latency}
\label{sec:cs3}

So far, we have assumed there is no extra latency between the local (near) memory as DRAM cache and the far (remote)
memory as its backing store. In this case study, we will analyze the impact of latency of the link between these two nodes
on the system's performance. For this purpose, we consider the system shown in Figure \ref{fig:case-studies} on the bottom right.
\begin{figure}

    \subfloat[DDR4]{%
      \includegraphics[clip,width=\linewidth]{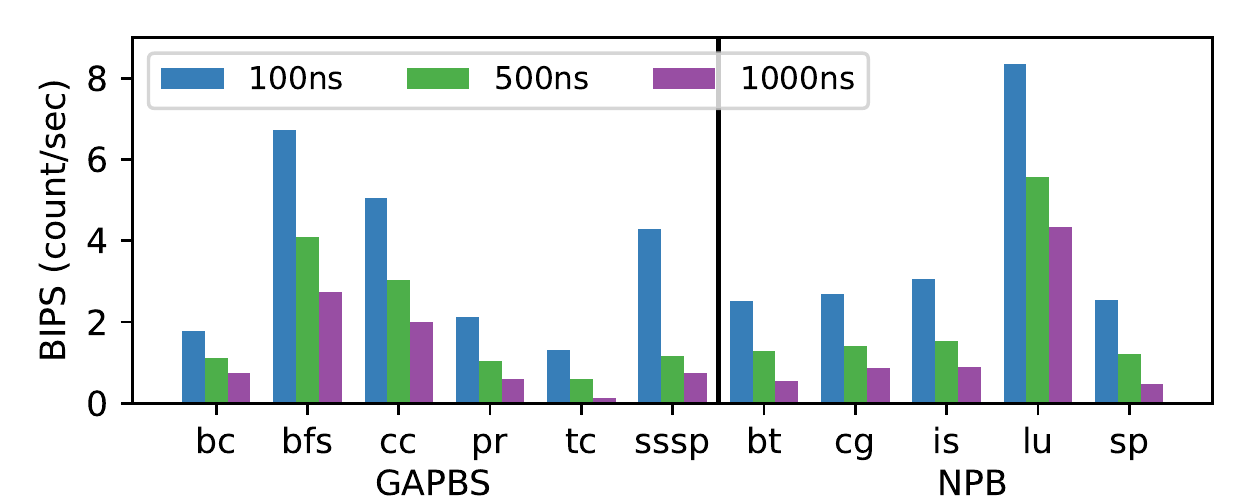}%
      \vspace{-1em}
      \label{fig:cs3BipsDdr4}
    }
    \vspace{-1em} \\
    \subfloat[NVM]{%
      \includegraphics[clip,width=\linewidth]{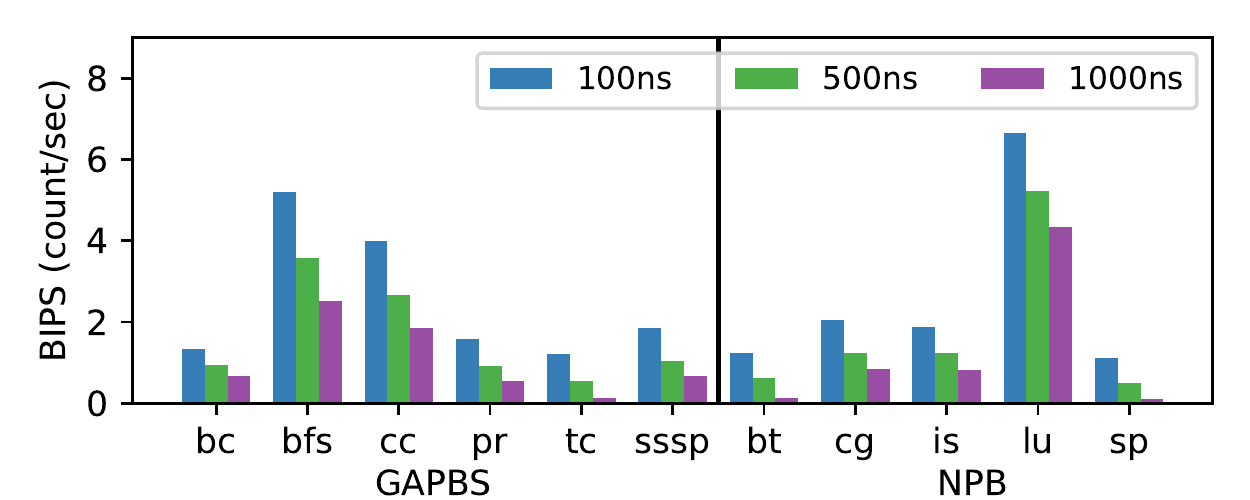}%
      \vspace{-1em}
      \label{fig:cs3BipsNvm}
    }
    \caption{Performance of GAPBS and NPB on DRAM cache for 100 ns, 500 ns, and 100 ns round-trip
    link latency of the far memory.
    BIPS refers to a billion instructions per second. DDR4 and NVM technologies are used for remote main memory.}
    \label{fig:cs3BipsDDR4NVM}
    \vspace{-2em}
\end{figure}



The DRAM cache model we described in this paper is capable of employing any memory technologies that are modeled in \gem{},
including DDR3, DDR4, HBM1, HBM2, and NVM. In this study, we use the baseline DRAM cache design (inspired by Intel's Cascade Lake). We use a single channel of HBM2 (2 pseudo-channels) DRAM cache
with a single channel of DDR4 remote backing store. In a second case, we change the remote main memory to a single channel of NVM. The DDR4 and NVM models of
\gem{} provide 19.2 GB/s theoretical peak bandwidth per single channel, though NVM has higher read and write latencies than DDR4.
We also add a link between the DRAM cache manager and the remote backing store
with a configurable latency, as shown in Figure \ref{fig:case-studies} on the bottom right. We conduct a full-system simulation to run GAPBS and NPB on the two systems in three
different cases where the round-trip latency of the link will be set to 100 ns, 500 ns, and 1000 ns.
Similar upper bound latency numbers are reported for direct attached, serially attached, and network attached memory devices in the related industrial~\cite{gupta2020genz} and academic~\cite{maruf2022tpp} work.
The rest of the methodology for the experiments remains the same as described in Section~\ref{method}.

\subsection*{Results and Discussions}

Figures \ref{fig:cs3BipsDdr4} and \ref{fig:cs3BipsNvm} show a billion instructions per second (BIPS) of DRAM cache systems while running GAPBS and NPB
workloads for DDR4 and NVM remote main memories, respectively.
For 100 ns link latency, the system with DDR4 performs better than the system with NVM.
This is expected given the higher read and write latencies of NVM devices compared to DDR4.

Once we increase the link latency to 500 ns, as shown in the figures, the DDR4 system still has higher performance compared to the NVM system.
However, the performance gap between the two systems in 500 ns is smaller than the gap in 100 ns link latency. 
Once the link latency increases to 1000 ns, for most of the workloads
the performance gap between DDR4 and NVM decreases compared to the 500 ns case.
As the link latency increases, the overall system performance drops in each system (DDR4 and NVM);
however, it drops less for the NVM system. 
With increased link latency both DDR4 and NVM remote memories have enough bandwidth to respond to the requests and the
latency of the NVM device gets amortized over the requests. 
Thus, the performance of the system with NVM drops less.

We also see in the figures that once the latency increase from 500 ns to 1000 ns the performances of $bt$ and $sp$
drop significantly in NVM system compared to DDR4 system.
These two workloads have the highest ratio of dirty line misses (Figure \ref{fig:cs1Mpki}). The DRAM
cache manager require to write the evicted dirty lines to the backing store. The limited write buffer of NVM
devices~\cite{wang2020characterizing} and the high link latency to access the remote NVM, create back pressure on the
NVM backing store. Thus, the performance drops significantly if the application is write intensive with a high miss
ratio for the NVM system.

\begin{figure}
    \center
    \subfloat[\scriptsize{DDR4}]{
    \includegraphics[width=.49\linewidth]{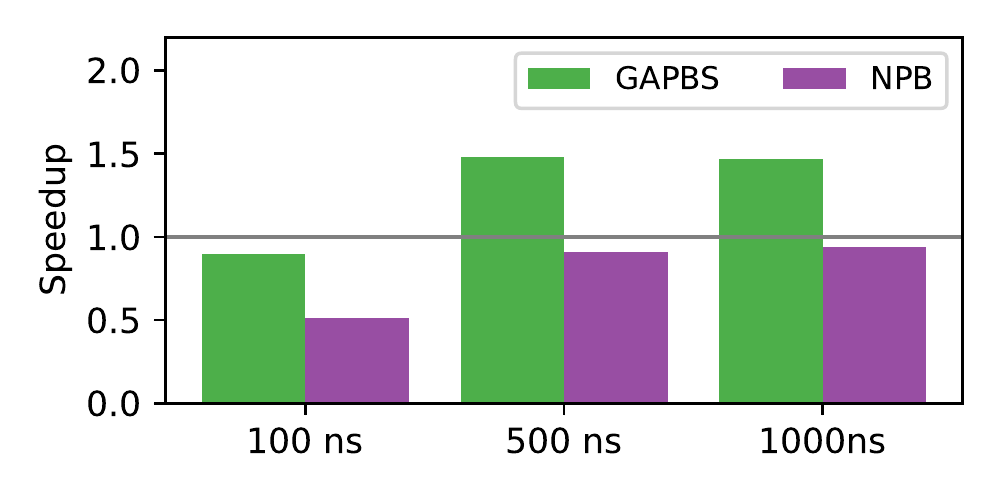}
    \label{fig:cs3SuDdr4}
    }
    \subfloat[\scriptsize{NVM}]{
    \includegraphics[width=.49\linewidth]{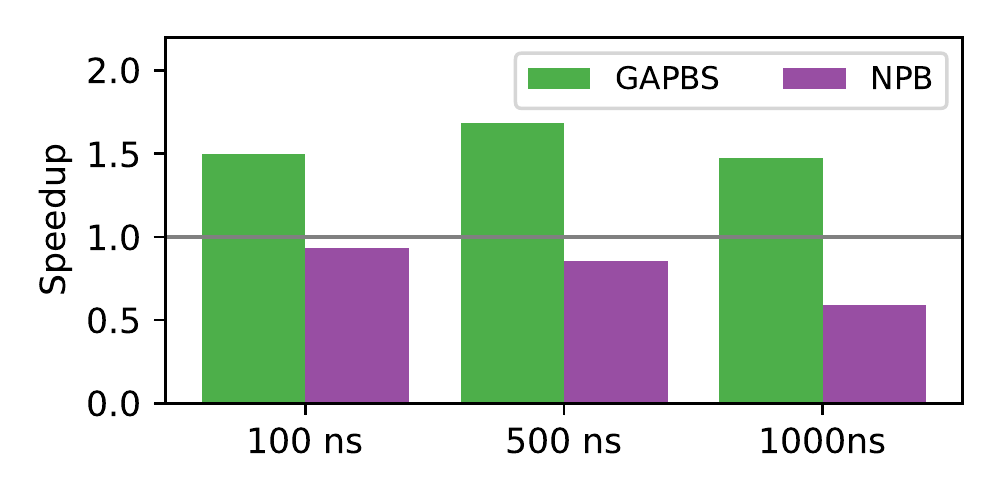}
    \label{fig:cs3SuNvm}
    }
    \caption{Speedup of DRAM cache system backed up by (a) DDR4 remote memory and (b) NVM remote memory compared to the same system without the DRAM cache.
    Geometric mean of workloads' throughput is used for speedup.}
    \label{fig:cs3SuDcacheMm}
    \vspace{-1.5em}
\end{figure}

Finally, we compare the performance of the DRAM cache systems to the same systems without the DRAM cache. Figures
\ref{fig:cs3SuDdr4} and \ref{fig:cs3SuNvm} show the speedup of the DRAM cache system compared to the same system
without the DRAM cache for DDR4 and NVM remote main memories, respectively. 
For GAPBS, Figure \ref{fig:cs3SuDdr4} shows once the link latency increases toward 500 ns and 1000 ns, the DRAM cache with DDR4 far main memory outperforms
the same system without the DRAM cache. However, as we observe in Figure \ref{fig:cs3SuNvm}, the DRAM cache with remote NVM main memory outperforms the same system without the DRAM cache for all the link latencies. The
differences between the read and write latencies of the DDR4 and NVM devices can explain the case for 100 ns link latency.
For NPB, the DRAM cache systems do not outperform the systems without the DRAM cache, for both DDR4 and NVM cases.
In Figures \ref{fig:cs1MissRatio} and \ref{fig:cs1Mpki} we observe that NPB is more write-intensive and has higher miss ratio compared to the GAPBS.
Thus, workloads in NPB generate more write backs to the main memory compared to GAPBS. Thus,
as the link latency grows, the pressure on the remote main memory increases. This becomes more critical for
NVM main memory systems due to their limited write buffer. This pressure does not exist in the system without DRAM cache.
Thus, NPB's performance drops as link latency increases in Figure \ref{fig:cs3SuNvm}.

\textbf{Takeaway:} This case study shows for systems with long latencies (e.g., 1000 ns) between local DRAM caches and remote main memories, NVM can perform close to
DDR4 devices. This can turn into an interesting use case to utilize large capacity of NVM devices. The study
also shows that for the workloads that are not write-intensive with high miss ratio, DRAM cache systems can outperform
the same system without the DRAM cache, if the remote main memory's link latency is higher than 100 ns.



\section{Related work}
\label{sec:related}

Today's high-performance computing systems (in HPC or cloud centers) under-utilize memory resources because of over-provisioning~\cite{peng2020memory,shalf2020photonic}.
Future HPC systems will decouple compute and memory extensively, thus leading to disaggregated architectures, where multiple processing elements can potentially access a large pool of memory resources using a coherent interconnect fabric (e.g., CXL~\cite{van2019hoti}, Gen-Z~\cite{casey2017gen})~\cite{awadascr21}.
Locally attached memories to a CPU can act as a cache to the remote pool of memory.
Given the promise of memory pooling, there is an increased interest in research on disaggregated memory systems recently~\cite{wahlgren2022evaluating, maruf2022tpp, calciu2021rethinking, kommareddy2021deact}.
We anticipate more research in this direction in the near future.
Our work provides the appropriate flexibility to model and evaluate the previously mentioned systems. The evaluation methods and tools adopted by previous works to evaluate DRAM cache designs are generally private. The lack of such tools is one of the main motivations for our work to build a flexible simulation model for DRAM caches which can be a part of a widely used open-source simulator like \gem{}.

The evaluation strategies discussed in prior DRAM cache research~\cite{qureshi2012fundamental, jevdjic2013stacked, chou2015bear, young2018accord, jevdjic2014unison} lack in many ways compared to our simulation model.
Many previous research works on DRAM cache techniques use trace-based or functional-first simulators.
These simulators might not faithfully simulate the behavior of applications that take different paths depending on I/O or thread timings~\cite{eeckhout2010computer}.
Trace-based simulators cannot model the mis-speculated execution path and the micro-architectural impact on thread order.
Functional-first simulators allow the functional model to execute ahead of the timing model, and thus, to model mispredicted path, they need to rely on rollback of functional model state~\cite{eeckhout2010computer}.
However, the timing differences in the functional and the timing models can change the order in which the functional model acquires a lock and what is observed by the timing model~\cite{eeckhout2010computer}.
In contrast, execute-in-execute simulators like \gem{}, which functionally execute instructions at the simulated execute stage, can be more accurate.

Notably, most of the previous research on DRAM caches does not use full-system simulators and might fail to capture the OS impact. Work by Bin et al.~\cite{gao2022level} has shown that the OS kernel bottlenecks can degrade memory access latency on DRAM cache systems in disaggregated environments. We implement our DRAM cache simulation model in a full-system simulator and use full-system simulations for all the results presented in this paper.
Additionally, many previous works use multiple copies of the same benchmark to execute workload on multiple simulated cores, which might not capture the inter-thread dependency behavior exhibited by the real workloads on real systems. In contrast, we used multithreaded workloads to use all the simulated cores in this work.

In addition to simulation methodologies, some work exists on using analytical models for DRAM caches. For example, Gulur et al.~\cite{gulur2015comprehensive} presented an analytical performance model of different DRAM organizations. However, their work is agnostic to the micro-architectural and timing constraints of main memory technologies cooperating with DRAM cache,  and still leaves a gap for a full system DRAM cache simulation for detailed architectural analysis.

\section{Conclusion}

In this work, we described our detailed cycle-level DRAM cache
model implemented in \gem{}, which enables design space exploration for DRAM caches
in emerging memory systems.
The model presented in this work
can enable many interesting research works in the domain of heterogeneous and disaggregated memory systems.
For instance, using our DRAM cache model we can address questions such as: what is
the efficient data placement and data movement policy and mechanism in systems composed of fast and slow memories.
Since our model is implemented in a full system simulation platform, it can also enable the hardware and software co-design
research in such systems.



\bibliographystyle{IEEEtran}
\bibliography{IEEEabrv,references}

\end{document}